\documentclass[preprints,article,accept,moreauthors,pdftex]{Definitions/mdpi} 

\firstpage{1} 
\pubvolume{00}
\issuenum{0}
\articlenumber{0}
\pubyear{2020}
\copyrightyear{2020}
\history{Received: 31 March 2020; Accepted: 15 May 2020; Published: date}
\updates{yes} 

\setitemize{parsep=6pt,itemsep=0pt,leftmargin=*,labelsep=5.5mm}
\setenumerate{parsep=6pt,itemsep=0pt,leftmargin=*,labelsep=5.5mm}
\setlist[description]{itemsep=0mm} 

\Title{Measurement-Based Adaptation Protocol with Quantum Reinforcement Learning in a Rigetti Quantum Computer}

\Author{Julio Olivares-S\'anchez $^{1}$, Jorge Casanova $^{1,2}$, Enrique Solano $^{1,2,3,4}$ and Lucas Lamata $^{1,5,}$*}%

\AuthorNames{Julio Olivares-S\'anchez, Jorge Casanova, Enrique Solano and Lucas Lamata}

\address{%
$^{1}$ \quad Department of Physical Chemistry, University of the Basque Country UPV/EHU, Apartado 644, \newline 48080 Bilbao, Spain; juliolivaresanchez@gmail.com (J.O.-S.); jcasanovamar@gmail.com (J.C.); enr.solano@gmail.com (E.S.)\\
$^{2}$ \quad IKERBASQUE, Basque Foundation for Science, Maria Diaz de Haro 3, 48013 Bilbao, Spain\\
$^{3}$ \quad International Center of Quantum Artificial Intelligence for Science and Technology (QuArtist)\\ and Physics Department, Shanghai University, Shanghai 200444, China\\
$^{4}$ \quad IQM, 
 Munich, Germany\\
$^{5}$ \quad Departamento de F\'isica At\'omica, Molecular y Nuclear, Universidad de Sevilla, 41080 Sevilla, Spain}

\corres{Correspondence: llamata@us.es}

\abstract{We present an experimental realisation of a measurement-based adaptation protocol with quantum reinforcement learning in a Rigetti cloud quantum computer. The experiment in this few-qubit superconducting chip faithfully reproduces the theoretical proposal, setting the first steps towards a semiautonomous quantum agent. This experiment paves the way towards quantum reinforcement learning with superconducting circuits.}


\begin{document}


\section{Introduction}

\vspace{0.5cm}

Quantum machine learning \cite{Measurement based adaptation protocol, key-2bis,key-2,key-3,key-4,key-5,key-6,key-7,key-8,key-12,key-13,key-14,LamataReview,key-15,key-16,key-17,key-18,key-19,key-20,key-22,key-23,key-24,Melnikov,key-25,key-28,key-26,key-27,key-1,key-29} is a field of research that has raised much attention in the past few years, especially for the expectation that it may enhance the machine learning calculations in current  and future technology. The machine learning field, inside artificial intelligence, is divided into three main areas: supervised learning, unsupervised learning    and reinforcement learning \cite{Russel Norvig}. The first two rely on training the system via labelled or unlabelled data, respectively. Moreover, the third one considers an intelligent agent that interacts with its outer world, the environment, gathering information from it, as well as acting on it, being employed, e.g., in robotics. In each learning iteration, the agent decides a strategy, or policy, on the best action to take, depending on its past history and goal oriented. Reinforcement learning can be considered as the most similar way in which human beings learn, via interactions with their outer world. 

Among the protocols being developed in quantum machine learning, the ones based on quantum reinforcement learning will enable the future deployment of semiautonomous quantum agents, which may produce significant advances in quantum computation and artificial intelligence \cite{key-2bis,key-25,key-26,key-27,Measurement based adaptation protocol,key-1,key-29}. Already some possible speedups when considering quantum agents have been suggested \cite{key-25}, while other interesting aspects of quantum reinforcement learning are related to quantum systems autonomously learning quantum data \cite{Measurement based adaptation protocol}.

In this article, we implement a protocol for measurement-based adaptation with quantum reinforcement learning, proposed in   \cite{Measurement based adaptation protocol}, in an 8-qubit cloud quantum computer provided by Rigetti. The experimental results have a good agreement with the theoretical expectations, establishing a plausible avenue for the future development of semiautonomous quantum agents. We point out that further efforts in experimental quantum machine learning, both in superconducting circuits \cite{RistenpjQuInfo} and in quantum photonics \cite{Xanadu}, are being pursued.

\section{Results}
\vspace{-6pt}

\subsection{Measurement-Based Adaptation Protocol with Quantum Reinforcement
Learning\label{sec:Measurement-based-adaptation-pro}}

We review the protocol of     \cite{Measurement based adaptation protocol}, which we subsequently implement in the Rigetti cloud quantum computer.
 The aim of this algorithm is to adapt a quantum state to a reference
unknown state via successive measurements. Many identical copies of
the reference state are needed for this protocol and, after each measurement, which destroys the state, more information about it is obtained. The
system that we consider is composed of the following parts:
\begin{enumerate}
\item The \emph{environment} system  (E) contains the reference state copies.
\item The \emph{register} (R) interacts with E and obtains information
from it.
\item The \emph{agent} (A) is adapted by digital feedback depending on the
outcome of the measurement of the register.
\end{enumerate}

Let us assume that we know the state of a quantum system called agent
and that many copies of an unknown quantum state called environment
are provided. Let us also consider an auxiliary system called register
that interacts with E. Thus, we extract information from E by
measuring R and use the result as an input for a reward function (RF).

Subsequently, we perform a partially-random unitary transformation
on A, which depends on the output of the RF.

Let us present the simplest case in which each subsystem is described
by a qubit state: Agent $|0\rangle_{A}$, Register $|0\rangle_{R}$ and
Environment $|\varepsilon\rangle_{E}=\left[\cos\left(\frac{\theta^{(1)}}{2}\right)|0\rangle_{E}+e^{i\phi^{(1)}}\sin\left(\frac{\theta^{(1)}}{2}\right)|1\rangle_{E}\right]$.

Therefore, the initial state is,
\begin{equation}
|\psi_{0}^{(1)}\rangle=|0\rangle_{A}|0\rangle_{R}\left[\cos\left(\frac{\theta^{(1)}}{2}\right)|0\rangle_{E}+e^{i\phi^{(1)}}\sin\left(\frac{\theta^{(1)}}{2}\right)|1\rangle_{E}\right].\label{eq:initial state}
\end{equation}

Then, we apply a CNOT gate with E as control and R as target (policy)  to obtain information from E, namely
\begin{equation}
{\rm C_E NOT_{R}}|\psi_{0}^{(1)}\rangle=|0\rangle_{A}\left[\cos\left(\frac{\theta^{(1)}}{2}\right)|0\rangle_{R}|0\rangle_{E}+e^{i\phi^{(1)}}\sin\left(\frac{\theta^{(1)}}{2}\right)|1\rangle_{R}|1\rangle_{E}\right].
\end{equation}

Secondly, we measure the register qubit in the basis $\left\{ |0\rangle,\,|1\rangle\right\} $
with probability $P_{0}^{(1)}=\cos^{2}\left(\frac{\theta^{(1)}}{2}\right),$
or $P_{1}^{(1)}=\sin^{2}\left(\frac{\theta^{(1)}}{2}\right)$ of obtaining
the state $|0\rangle$ or $|1\rangle$, respectively. Depending on the
result of the measurement, we either do nothing if the result is $|0\rangle$,
since it means that we collapse E into A, or perform a partially-random
unitary operator on A, if the result is $|1\rangle.$ This unitary transformation
(action) is given by

\begin{equation}
U_{A}^{(1)}\left(\alpha^{(1)},\beta^{(1)}\right)=e^{-iS_{A}^{Z(1)}\alpha^{(1)}}e^{-iS_{A}^{X(1)}\beta^{(1)}},
\end{equation}
 where $S^{k(1)}_A$ is the $k${th} spin component; $\alpha^{(1)}=\xi_{\alpha}\Delta^{(1)}$
and $\beta^{(1)}=\xi_{\beta}\Delta^{(1)}$ are random angles; and $\xi_{\alpha},\,\xi_{\beta}\in\left[-\frac{1}{2},\,\frac{1}{2}\right]$ are
random numbers. The range of the random numbers is $\alpha^{(1)},\,\beta^{(1)}\in\left[-\frac{\Delta^{(1)}}{2},\,\frac{\Delta^{(1)}}{2}\right]$.

Now, we initialise the register qubit state and use a new copy of
the environment obtaining the following initial state for the second
iteration,
\begin{equation}
|\psi_{0}^{(2)}\rangle=\mathcal{U}_{A}^{(1)}|0\rangle_{A}|0\rangle_{R}|\varepsilon\rangle_{E},
\end{equation}
 with $\mathcal{U}_{A}^{(1)}=\left[m^{(1)}U_{A}^{(1)}\left(\alpha^{(1)},\beta^{(1)}\right)+\left(1-m^{(1)}\right)\mathbb{I}_{A}\right]$
where $m^{(1)}=\left\{ 0,\,1\right\} $ is the outcome of the measurement,
and we call $\mathcal{U}_{A}^{(1)}|0\rangle_{A}\equiv|\overline{0}\rangle_{A}^{(2)}$.

Later, we define the RF as
\begin{equation}
\Delta^{\left(k\right)}=\left[(1-m^{(k-1)})\mathcal{R}+m^{(k-1)}\mathcal{P}\right]\Delta^{\left(k-1\right)}.
\end{equation}

 As we can see, the exploration range of the $k${th} iteration,
$\Delta^{\left(k\right)}$, is modified by $\mathcal{R}\Delta$ when
the outcome of the $k-1$ iteration was $m^{\left(k-1\right)}=0$
and by $\mathcal{P}\Delta$, when it was $m^{\left(k-1\right)}=1$. Here,  we   chose  $\mathcal{R}=\epsilon<1$, $\mathcal{P}=\frac{1}{\epsilon}$, with $\epsilon$ being a constant.

We define the value function (VF) in this protocol as the value of
$\Delta^{\left(n\right)}$ after many iterations assuming that $\Delta\rightarrow0$,
i.e., that the agent converges to the environment state.

In the $k${th} iteration of the protocol, we assume that the system starts in the following state,
\begin{equation}
|\psi\rangle^{\left(k\right)}=|\overline{0}\rangle_{A}^{(k)}|0\rangle_{R}|\overline{\varepsilon}\rangle_{E},
\end{equation}
 where $|\overline{0}\rangle_{A}^{\left(k\right)}=\text{\ensuremath{\mathbb{U}}}^{\left(k\right)}|0\rangle_{A}$,
$|\overline{\varepsilon}\rangle_{E}=\mathbb{U}^{\left(k\right)\dagger}|\varepsilon\rangle_{E}$,
and the accumulated rotation operator is,
\begin{equation}
\text{\ensuremath{\mathbb{U}}}^{\left(k\right)}=\left[m^{(k-1)}U^{(k-1)}\left(\alpha^{(k-1)},\beta^{(k-1)}\right)+\left(1-m^{(k-1)}\right)\mathbb{I}\right]\text{\ensuremath{\mathbb{U}}}^{\left(k-1\right)},
\end{equation}
 with $\text{\ensuremath{\mathbb{U}}}^{\left(1\right)}=\mathbb{I}$.
 
Then, we perform the gate ${\rm C_{E}NOT_{R}}$,
\begin{equation}
|\phi\rangle^{\left(k\right)}={\rm C_{E}NOT_{R}}|\overline{0}\rangle_{A}^{(k)}|0\rangle_{R}|\overline{\varepsilon}\rangle_{E}=|\overline{0}\rangle_{A}^{(k)}\left[\cos\left(\frac{\theta^{\left(k\right)}}{2}\right)|0\rangle_{R}|0\rangle_{E}+e^{i\phi^{\left(k\right)}}\sin\left(\frac{\theta^{\left(k\right)}}{2}\right)|1\rangle_{R}|1\rangle_{E}\right],
\end{equation}
 and measure $R$, with probabilities $P_{0}^{\left(k\right)}=\cos^{2}\left(\frac{\theta^{\left(k\right)}}{2}\right)$
and $P_{1}^{\left(k\right)}=\sin^{2}\left(\frac{\theta^{\left(k\right)}}{2}\right)$,
for the outcomes $m^{\left(k\right)}=0$ and $m^{\left(k\right)}=1$,
respectively. Finally, we update the reward function, $\Delta^{\left(k\right)}$, for the next iteration.

\subsection{Experimental Setup: Rigetti Forest Cloud Quantum Computer\label{part:Experimental-setup}}

Cloud quantum computers are dedicated quantum processors operated
by users logged through the Internet (cloud). Although D-Wave Systems, Inc. \cite{DWave}
was among the first companies to commercialise quantum computers in 2011,
it was not until the arrival of IBM Quantum Experience \cite{IBMQExperience} in May 2016
that there was a quantum computer openly available in the cloud. Approximately one
year later, in June 2017, Californian company Rigetti Computing \cite{Rigetti} announced
the availability of a cloud quantum computing platform. This last quantum platform is the one that we used in this work, because of convenience of their web interface to implement our protocol with feedback. In the past year, Rigetti built a brand
new processor of 8 qubits called 8Q-Agave: this is the chip we employed. An advantage of this device is that one does not
have to adapt the algorithm to the topology of the
system. The compiler does it for us. In particular, it is the possibility of defining quantum gates in
matrix form that makes the adaptation significantly simpler. Among other features, Rigetti Forest offers the possibility of simulating
the experiments in their quantum virtual machine (QVM). 

\subsubsection{Python-Implemented Algorithm}

In this section, we   explain how the algorithm of Section \ref{sec:Measurement-based-adaptation-pro}
is adapted to be implemented in Rigetti  simulator and quantum processor.
Firstly, we must initialise some variables and constants to correctly perform the first iteration,
\begin{itemize}
\item Reward and punishment ratios: $\epsilon\in\left(0,\,1\right)\Rightarrow\mathcal{R}=\epsilon$
and $\mathcal{P}=1/\epsilon$.
\item Exploration range: $\Delta=4\pi$.
\item The unitary transformation matrices: $\mathcal{U}=\mathbb{U}=\mathbb{U^{\dagger}}=\left(\begin{array}{cc}
1 & 0\\
0 & 1
\end{array}\right)$.
\item Partially-random unitary operator: $U\left(x,\,y\right)=\left(\begin{array}{cc}
e^{-i\frac{x}{2}} & 0\\
0 & e^{i\frac{x}{2}}
\end{array}\right)\left(\begin{array}{cc}
\cos\frac{y}{2} & -i\sin\frac{y}{2}\\
-i\sin\frac{y}{2} & \cos\frac{y}{2}
\end{array}\right)$.
\item Initial values of the random angles: $\alpha=\beta=0$. Makes $U\left(\alpha,\,\beta\right)=\mathbb{I}_{2}$
for the first iteration.
\item Initial value of the iteration index: $k=1$.
\item Number of iterations: $N$.
\end{itemize}

The algorithm is composed of the following steps,
\begin{enumerate}
\item Step 1: While $k<N+1\Rightarrow$, go to  Step 2.
\item Step 2: If $k\ne1\Rightarrow$
\begin{eqnarray}
&&\xi_{\alpha}=random\,number\in\left[-\frac{1}{2},\,\frac{1}{2}\right]\\\nonumber
&&\xi_{\beta}=random\,number\in\left[-\frac{1}{2},\,\frac{1}{2}\right],\label{eq:random numbers}
\end{eqnarray}
\begin{equation}
\begin{array}{c}
\alpha=\xi_{\alpha}\Delta\\
\beta=\xi_{\beta}\Delta
\end{array},\label{eq:random angles}
\end{equation}
\begin{equation}
\mathcal{U}=\left[m U\left(\alpha,\,\beta\right)+\left(1-m\right)\mathbb{I}_{2}\right],\label{eq:U_cal}
\end{equation}
\begin{equation}
\mathbb{U}=\mathcal{U}\cdot\mathbb{U},\label{eq:U_bb}
\end{equation}
\begin{equation}
\mathbb{U}^{\dagger}=\left(\mathbb{U}^{*}\right)^{T}.\label{eq:U_bb dagger}
\end{equation}
\item Step 3: First quantum algorithm.\\
First, we define the agent, environment and register qubits as,
\begin{equation}
|0\rangle_{A}|0\rangle_{R}|0\rangle_{E},\label{eq:qubit definition}
\end{equation}
and act upon the environment,
\begin{equation}
|\varepsilon\rangle_{E}=U^{E}|0\rangle_{E}.\label{eq:Environment}
\end{equation}
 Then, we have 
\begin{equation}
\mathbb{U}^{\dagger}|\varepsilon\rangle_{E}=|\overline{\varepsilon}\rangle_{E}.\label{eq:Ubbd sobre environment}
\end{equation}
We apply the policy
\begin{equation}
{\rm C_{E}NOT_{R}}|0\rangle_{R}|\overline{\varepsilon}\rangle_{E},\label{eq:policy}
\end{equation}
and measure the register qubit storing the result in $m=\left\{ 0,\,1\right\} $.
\item Step 4: Second quantum algorithm.\\
Subsequently, we act with $\mathbb{U}$ on the agent qubit in order to approach it to the environment state, $|\varepsilon\rangle_{E}$:
\begin{equation}
\mathbb{U}|0\rangle_{A},\label{eq:Ubb sobre Agent}
\end{equation}
Afterwards, we measure this qubit and store the result in a classical
register array. We repeat Step 4 a total   of     8192    times  to determine the state created after applying $\mathbb{U}$. 
\item In this last step, we apply the reward function,
\begin{equation}
\Delta=\left[\left(1-m\right)\mathcal{R}+m\mathcal{P}\right]\Delta,\label{eq:Reward function}
\end{equation}
and increase the iteration index by one after it: $k=k+1$. Go to
Step 1.
\end{enumerate}

\subsection{Experimental Results of Quantum Reinforcement Learning with the Rigetti Cloud
Quantum Computer\label{part:Results}}

In this section, we describe the experimental results with the Rigetti cloud quantum computer of the measurement-based adaptation protocol of     \cite{Measurement based adaptation protocol}.

The algorithm has been proved for six different initial states of the
environment. The states contain different weights of the quantum superpositions of $|0\rangle$ and $|1\rangle$ states, as well as different complex relative phases, to illustrate the performance of the protocol in a variety of situations. These states are the
following,
\begin{equation}
|\varepsilon_{1}\rangle_{E}=R_{Z}\left(\frac{\pi}{3}\right)R_{Y}\left(\frac{4\pi}{9}\right)|0\rangle_{E}=\left(\begin{array}{c}
e^{-i\frac{\pi}{6}}\cos\left(\frac{2\pi}{9}\right)\\
e^{i\frac{\pi}{6}}\sin\left(\frac{2\pi}{9}\right)
\end{array}\right)\approx\sqrt{0.6}|0\rangle+\sqrt{0.4}e^{i\frac{\pi}{3}}|1\rangle\label{eq:state 1},
\end{equation}

\begin{equation}
|\varepsilon_{2}\rangle_{E}=R_{Z}\left(\frac{\pi}{4}\right)R_{Y}\left(\frac{5\pi}{9}\right)|0\rangle_{E}=\left(\begin{array}{c}
e^{-i\frac{\pi}{8}}\cos\left(\frac{5\pi}{18}\right)\\
e^{i\frac{\pi}{8}}\sin\left(\frac{5\pi}{18}\right)
\end{array}\right)\approx\sqrt{0.4}|0\rangle+\sqrt{0.6}e^{i\frac{\pi}{4}}|1\rangle, \label{eq:state 2 Env}
\end{equation}
\begin{equation}
|\varepsilon_{3}\rangle_{E}=R_{Y}\left(\frac{\pi}{3}\right)|0\rangle_{E}=\left(\begin{array}{c}
\cos\left(\pi/6\right)\\
\sin\left(\pi/6\right)
\end{array}\right)=\sqrt{0.75}|0\rangle+\sqrt{0.25}|1\rangle, \label{eq:Environement state 3}
\end{equation}

\begin{equation}
|\varepsilon_{4}\rangle_{E}=R_{Y}\left(\frac{2\pi}{3}\right)|0\rangle_{E}=\left(\begin{array}{c}
\cos\left(\pi/3\right)\\
\sin\left(\pi/3\right)
\end{array}\right)=\sqrt{0.25}|0\rangle+\sqrt{0.75}|1\rangle, \label{eq:environment state 4}
\end{equation}
\begin{equation}
|\varepsilon_{5}\rangle_{E}=R_{Z}\left(\frac{\pi}{2}\right)H|0\rangle_{E}=\frac{1}{\sqrt{2}}\left(|0\rangle+i|1\rangle\right), \label{state 5}
\end{equation}

\begin{equation}
|\varepsilon_{6}\rangle_{E}=H|0\rangle_{E}=|+\rangle_{E}=\frac{1}{\sqrt{2}}\left(|0\rangle+|1\rangle\right). \label{eq:state |+> Environment}
\end{equation}

Hereunder, we plot the exploration range, $\Delta$,     and the fidelity, which we define in Equation     (\ref{ClasFidel}),
in terms of the total number of iterations. The results obtained using
the 8Q-Agave chip of Rigetti are shown with the corresponding ideal
simulation with the Rigetti quantum virtual machine,  which includes no noise. We point out that each of the experimental plots in this article contains a single realisation of the experiment, given that, in the adaptation protocol, single instances are what are relevant instead of averages. This is due to the fact that the successive measurements influence the subsequent ones and are influenced by the previous ones, in each experiment, in order for the agent to adapt to the environment. This has as a consequence the fact that the theory curve and the experimental curve for each example match qualitatively but not totally quantitatively, as they are both probabilistic. On the other hand, when the exploration range converges to zero we always observe convergence to a large final fidelity. The blue solid line represents the real experiment result while the red dashed-dotted
one corresponds to the ideal simulation. It is also worth mentioning
that after doing many experiments with the ideal and real simulators
for different values of the parameter $\epsilon$, we fixed it to
$\epsilon=0.95$. We found it to yield a balanced exploration--exploitation
ratio. The exploration--exploitation balance is a feature of reinforcement learning, where the optimal strategy must deal with exploring enough new possibilities while at the same time exploiting the ones that work best \cite{Russel Norvig}.

\begin{figure}[H]
\centering
\includegraphics[scale=0.34]{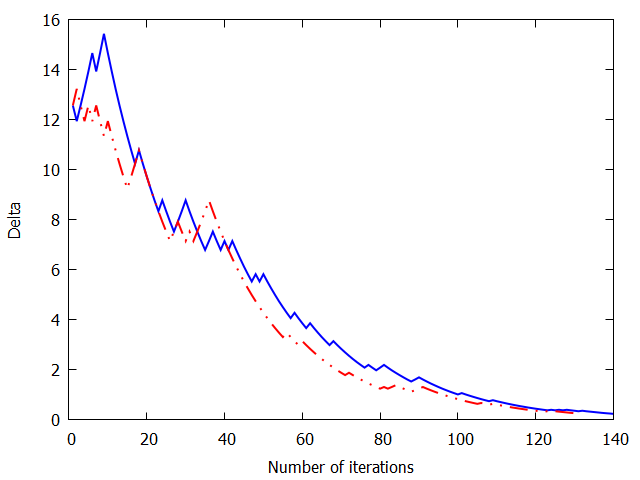}\includegraphics[scale=0.34]{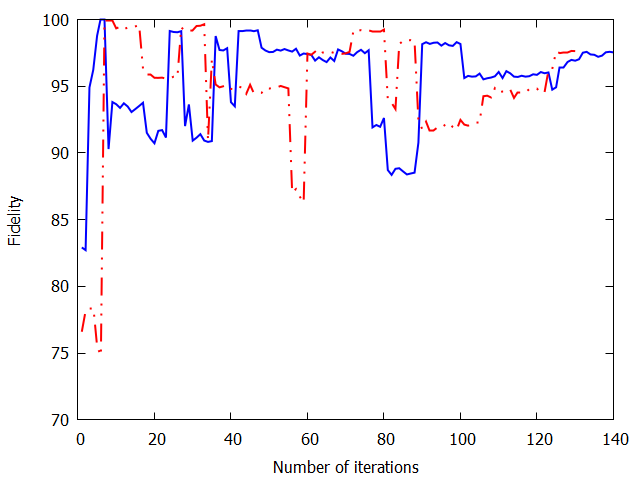}
\caption{Measurement-based adaptation protocol for the environment state $|\varepsilon_{1}\rangle_{E}$. The blue solid
line corresponds to the real experiment and the red dashed-dotted line
represents the ideal simulation. The fidelity is given in percentage (\%). \label{fig:MBAP-for-the1}}
\end{figure}

To begin with, let us have a look at Figure \ref{fig:MBAP-for-the1}, in which the environment state is $|\varepsilon_1\rangle_E$. It shows the exploration ratio, $\Delta,$ and the fidelity as defined in Equation (\ref{ClasFidel}) in terms of the number of iterations. As we can see,
140 iterations are enough for $\Delta$ to take a value of almost
zero. It is a good example of a balanced exploration versus exploitation
ratio, that is, the exploration ratio decreases making continuous
peaks. Each of them represents an exploration stage where $\Delta$
increases at the same time that the fidelity changes significantly.
These changes might bring a positive result, such that the agent receives
a reward, which means that $\Delta$ decreases, or a punishment and
it keeps increasing. Thus, the fidelity is not constant and it does
not attain a constant value of around 95\% until it has done 100 iterations.
In this case real and ideal experiments yield a similar result. It
is true that the ideal $\Delta$ decreases more smoothly and quickly  than
the real one. However, the values of the fidelity after 130 iterations
are practically the same.

In our calculations, we employ a classical measure of the fidelity. The reason to use the classical fidelity,     and not the quantum version,
is the reduction of needed resources in the experiments. We cannot make full quantum
state tomography as it is exponentially hard. Therefore, the definition
used in the algorithm is,
\begin{equation}
\mathcal{F}=\sqrt{p_{0}p_{0}^{T}}+\sqrt{p_{1}p_{1}^{T}}\label{ClasFidel}
\end{equation}
 for one-qubit measurements. Here, $p_{0}$ and $p_{1}$ stand for
the probability of obtaining $|0\rangle$ or $|1\rangle$ as an outcome when
measuring the real qubit and $p_{0}^{T}$ and $p_{1}^{T}$ are the
same probabilities for the corresponding theoretical qubit state that
we expect. This fidelity coincides at lowest order with the fully quantum one, illustrating the convergence of the protocol for a large number of iterations.

\begin{figure}[H]
\centering
\includegraphics[scale=0.34]{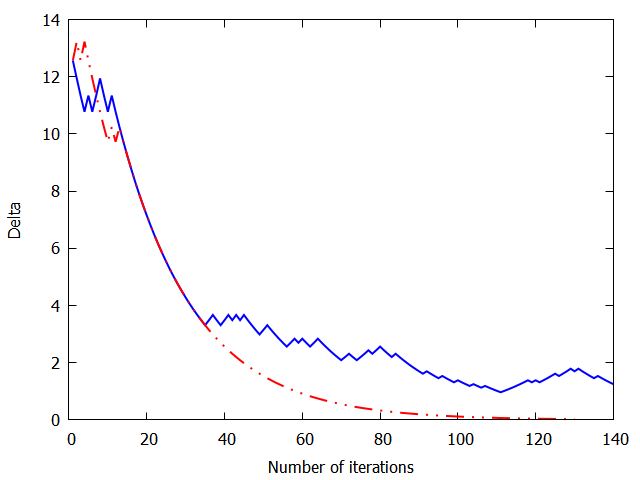}\includegraphics[scale=0.34]{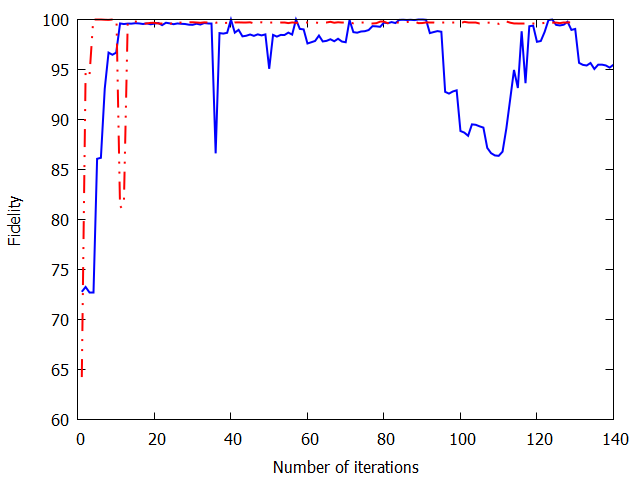}\llap{\shortstack{%
 \includegraphics[scale=0.18]{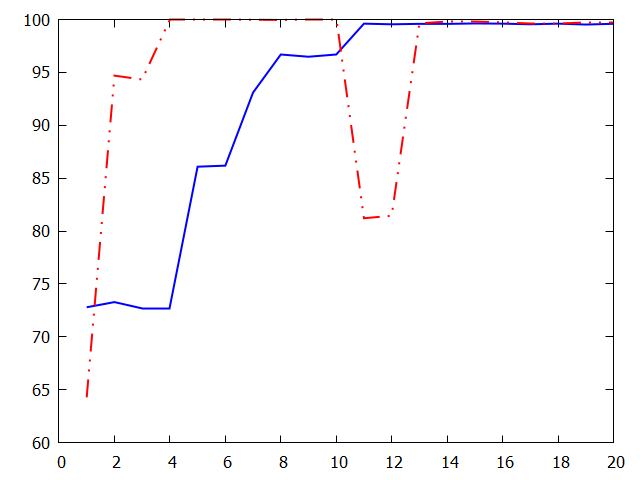}\\
 \rule{0ex}{0.4in}%
 }
 \rule{0.2in}{0ex}} \caption{Measurement-based adaptation protocol for the environment state $|\varepsilon_{2}\rangle_{E}$. The blue solid
line corresponds to the real experiment and the red dashed-dotted line
represents the ideal simulation. The fidelity is given in percentage (\%).\label{fig:MBAP-for-the2}}
\end{figure}

Let us continue with the discussion focussing on Figure \ref{fig:MBAP-for-the2}.
In this experiment, the algorithm has to take the agent from the $|0\rangle$
state to the environment state $|\varepsilon_2\rangle_E$ which is closer to one than to zero
(0.6 is the probability of getting $|1\rangle$ as an outcome when measuring
the environment). Bearing this in mind, it seems reasonable that 70
iterations are not enough for $\Delta$ to reach a value below 2, in
the real case. Apart from this,   despite   achieving a value above
99\% of fidelity in fewer than 20 iterations, the exploration still
continues.  Consequently,  the agent drops from its current state  to
one further from $|\varepsilon_{2}\rangle_{E}$.

In general, we notice a clear relationship between how smooth the
$\Delta$ line is and how constant the fidelity remains. Indeed, the
exploration ratio $\Delta$ decreases smoothly from fewer than 20 iterations
to fewer than 40. In this range, the fidelity does not change because
the agent is being rewarded. The price to pay for not exploring at
so early stages of the learning is that the convergence of the delta
is produced for a larger number of iterations than in other experiments.
After 140 iterations, we see that the convergence of $\Delta\rightarrow0$
is not guaranteed, namely, in the real experiment, it has a value above 1. Regarding the ideal simulation result, we draw the conclusion that
fewer than 20 iterations could be enough to converge to the environment
state with fidelity larger than 99.9\% and, what is more, remaining
on the same state until the exploration range has converged to
zero (see inset in Figure \ref{fig:MBAP-for-the2}).

\begin{figure}[H]
\centering
\includegraphics[scale=0.34]{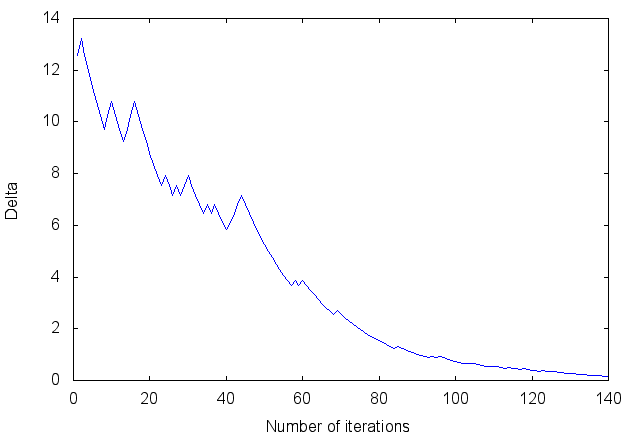}\includegraphics[scale=0.34]{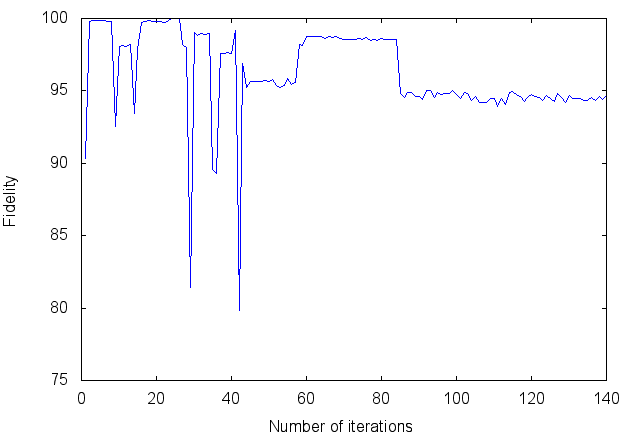}\caption{Measurement-based adaptation protocol for the environment state $|\varepsilon_{3}\rangle_{E}$. Experiment
one. The blue solid line corresponds to the real experiment. The fidelity is given in percentage (\%).\label{fig:---MBAPfor-the3-1}}
\end{figure}

\begin{figure}[H]
\centering
\includegraphics[scale=0.34]{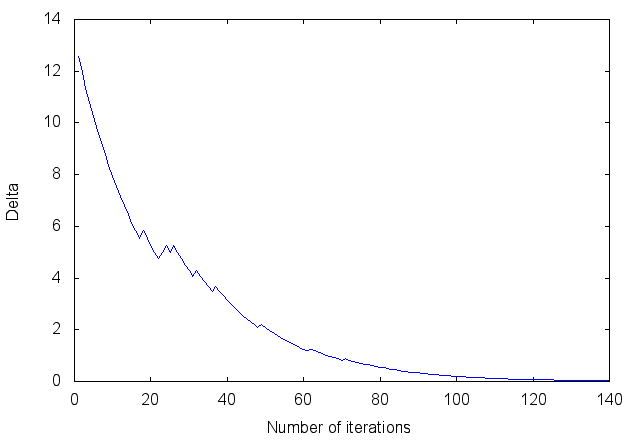}\includegraphics[scale=0.34]{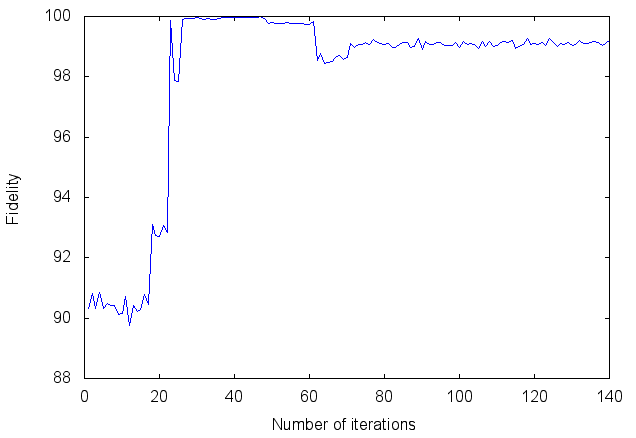}\llap{\shortstack{%
 \includegraphics[scale=0.18]{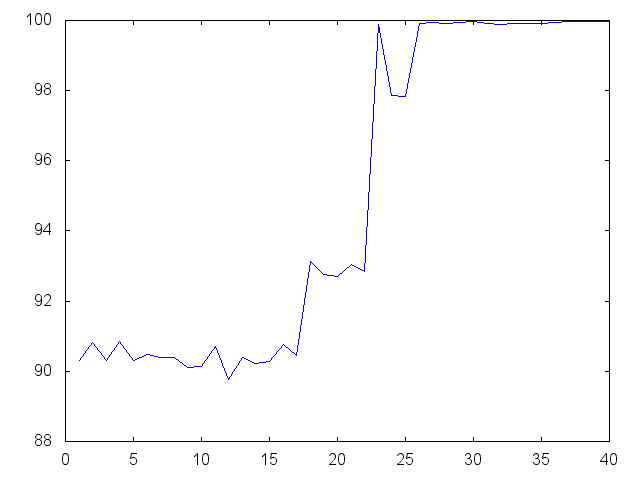}\\
 \rule{0ex}{0.4in}%
 }
 \rule{0.2in}{0ex}}

\caption{Measurement-based adaptation protocol for the environment state $|\varepsilon_{3}\rangle_{E}$. Experiment
two. The blue solid line corresponds to the real experiment. The fidelity is given in percentage (\%).\label{fig:--MBAP-for-the3-2}}
\end{figure}

In third place, we have the environment state $|\varepsilon_{3}\rangle_{E}$.
The results obtained using this initial state of the environment are
presented in Figures \ref{fig:---MBAPfor-the3-1} and \ref{fig:--MBAP-for-the3-2}.
Unlike the previous examples, we do not compare the ideal theory and real
experiments, which have similarly good agreement. Instead, we contrast two different real experiment outcomes.
In this way, we can show how even for the same experiment, i.e., initial
state of the environment, the algorithm can experience different behaviours.
In both cases, $\Delta$ goes to zero and the fidelity reaches a constant
value above 94\% in the first case and 99\% in the second one. However,
this convergence is achieved in two different ways. On the one hand,
we observe that exploitation predominates over exploration (see Figure \ref{fig:--MBAP-for-the3-2}),
except for several spots where the algorithm keeps exploring new
options. Then, as the initial fidelity is larger than 90\%, the state
of the agent converges to the environment with fewer than 70 iterations.
On the other hand, when exploration is more important (as shown in Figure \ref{fig:---MBAPfor-the3-1})
the fidelity is erratic, changing from low to high values. Moreover,
it takes longer for $\Delta$ to converge and for the fidelity to
be stabilised---more than 80 iterations.

Let us focus just on the first stages of the learning process, for
fewer than 40 iterations. Comparing both experiments, we see that, in
the first case, it starts exploring from the very beginning; thus,
with fewer than 20 iterations, the fidelity takes a value above 99.6\%.
   In the other case, there is more exploitation at the beginning
and around 25 iterations are required to reach a fidelity of 99\%. 

Among the six states that we have chosen, this is the one in which
agent and environment are the closest. Nevertheless, 70 iterations
are not enough to reach a value of $\Delta$ below 1. Thus, we can state
that a smaller distance in the Bloch sphere between agent and environment
does not imply in general a faster learning.

\begin{figure}[H]
\centering
\includegraphics[scale=0.34]{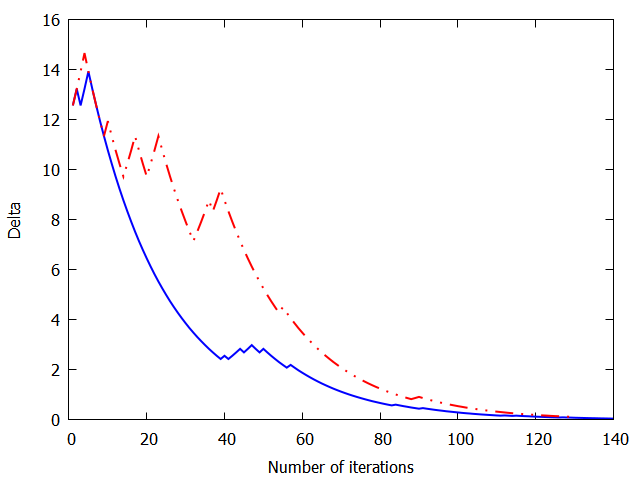}\includegraphics[scale=0.34]{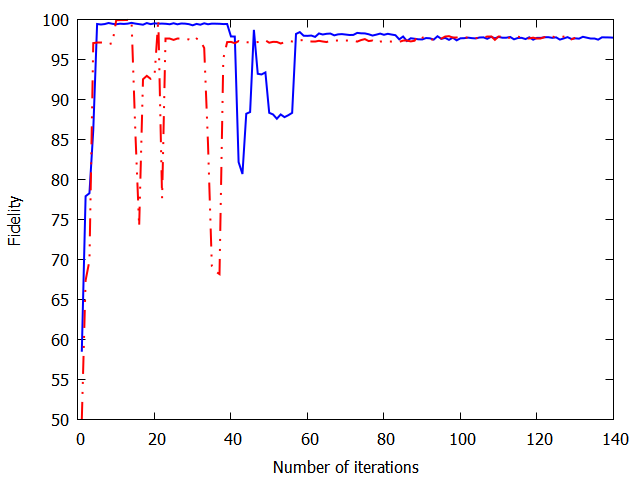}
\caption{Measurement-based adaptation protocol for the environment state $|\varepsilon_{4}\rangle_{E}$. The blue solid line corresponds to the real experiment and the red dashed-dotted line
represents the ideal simulation. The fidelity is given in percentage (\%).\label{fig:MBAP-for-the4}}
\end{figure}

Let us analyse now Figure \ref{fig:MBAP-for-the4}. This state is again
the most asymmetric case along with the previous one. However, unlike
the previous experiment, this one begins with the lowest value of  
fidelity. The environment state is the farthest one to the initial
agent state $|0\rangle_{A}$, with just a probability of 0.25 of achieving
this outcome (zero) when measuring the environment. Therefore, as
  might be expected, the algorithm is still exploring after 70 iterations
rather than exploiting the knowledge it has already acquired from
the environment. It is also proved that fewer than 100 iterations can
be enough to reach a value of $\Delta$ below 1. Then, in this case,
it is proved that the agent has already converged to a state with
fidelity larger than 97\%. It is also remarkable how, in this case,
with fewer than 10 iterations, the fidelity has attained a value larger
than 99\%. However, as the delta had not converged yet, it goes out
of this value later, exploring again. Once again, as a general rule, we can see that the algorithm is exploring for all the iterations. To explore is a synonym of changing fidelity, while, whenever the delta decreases smoothly, the fidelity remains constant. With this result, we wanted to show how sometimes the real experiment
converged more quickly, e.g., with just  nine iterations, to the environment state.
On the top of that, the exploration range also went   to zero more quickly 
than the ideal experiment. Nevertheless, the value of fidelity when
$\Delta$ has converged is exactly the same in both cases.

\begin{figure}[H]
\centering
\includegraphics[scale=0.34]{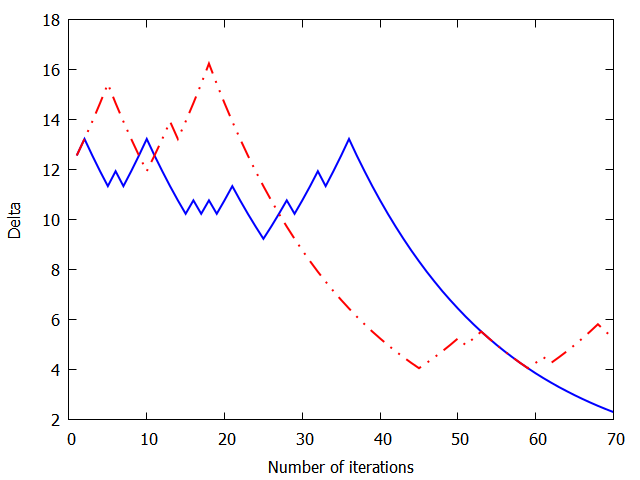}\includegraphics[scale=0.34]{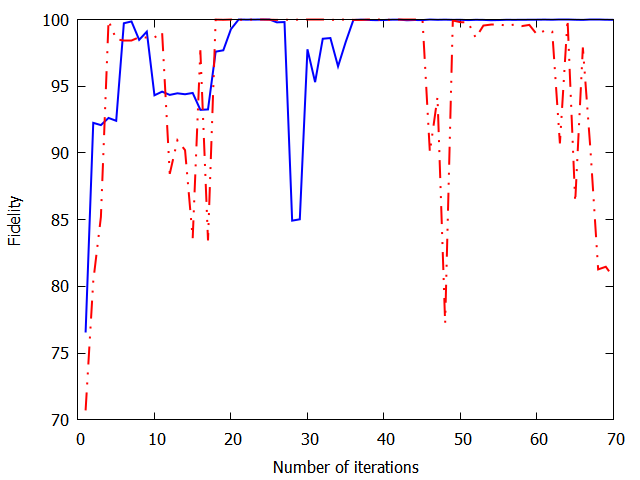}

\caption{Measurement-based adaptation protocol for the environment state $|\varepsilon_{5}\rangle_{E}$. The blue solid
line corresponds to the real experiment and the red dashed-dotted line
represents the ideal simulation. The fidelity is given in percentage (\%).\label{fig:MBAP-for-the5}}
\end{figure}

We analyse now the most symmetric cases, where the environment is prepared
in a uniform superposition with a relative phase between both states. The experiment chosen to highlight here (see Figure \ref{fig:MBAP-for-the5})
is the one in which the fidelity reaches a constant value above 99.9\%
in fewer than 40 iterations. The corresponding $\Delta$ evolves with
a good balance between exploration and exploitation until reaching
a point where it does not explore anymore and decreases very smoothly.
Comparing it to the ideal case, we notice two opposing behaviours:
 the ideal case makes a larger exploration at the beginning
which yields a larger constant value of the fidelity with fewer than
20 iterations, whereas the real system needs almost 40 iterations to get to this
point. On the other hand, in the real experiment, there is a larger
learning stage from 0 to 40 iterations. Thus, in this particular
case, the exploration ratio diminishes  more quickly  in the real experiment
for a large number of iterations. It happens because of the larger
value of $\Delta$ attained in the ideal experiment.

\begin{figure}[H]
\centering
\includegraphics[scale=0.34]{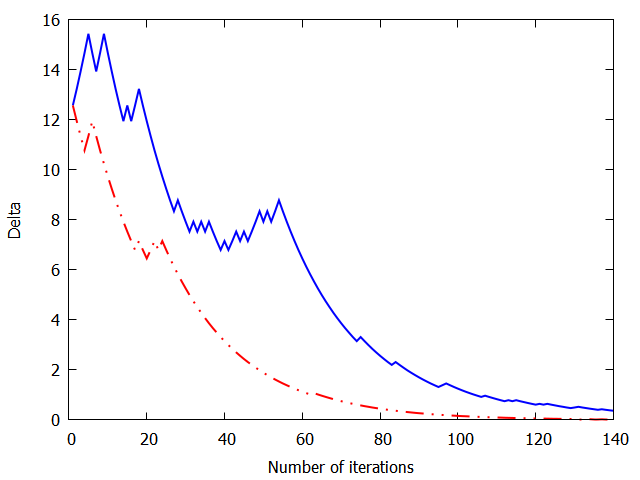}\includegraphics[scale=0.34]{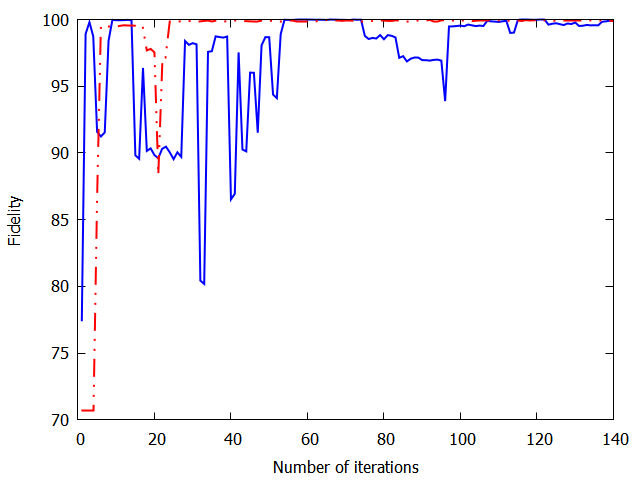}\llap{\shortstack{%
 \includegraphics[scale=0.18]{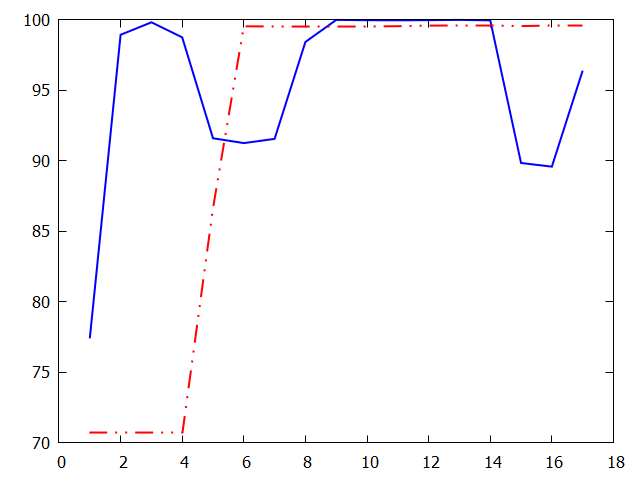}\\
 \rule{0ex}{0.4in}%
 }
 \rule{0.2in}{0ex}}

\caption{Measurement-based adaptation protocol for the environment state $|\varepsilon_{6}\rangle_{E}$. The blue solid
line corresponds to the real experiment and the red dashed-dotted line
represents the ideal simulation. The fidelity is given in percentage (\%).\label{fig:MBAP-for-the6}}
\end{figure}

Finally, in this second symmetric case $|\varepsilon_{6}\rangle_{E}$,
there are no relative phases between the states of the computational
basis. This experiment is very rich in phenomena and exemplifies very
well how the algorithm works. Figure \ref{fig:MBAP-for-the6} shows
clearly the fast increase of the fidelity until reaching values above
99.9\% with just nine iterations. Initially, there is an exploration
stage that makes the fidelity grow up to 99.9\% with just  nine  iterations.
At the same time, the exploration range, $\Delta$, grows, making two
consecutive peaks and then it decreases smoothly, while the fidelity
remains constant. Subsequently, there are a couple of exploration
peaks that make the fidelity oscillate. Now, after a few iterations
where the fidelity decreases smoothly, we come to a third and most
important exploration phase where we observe how the fidelity has
an increasing tendency. It suffices to check that the subsequent minima
of the fidelity take larger and larger values. Such an amount of exploration
has as a long-term reward a fidelity of 99.99\% after fewer than 70
iterations. However, the exploration range is still large and leaves
room for trying new states which make the fidelity drop again. Finally, the algorithm is able to find a good exploration-exploitation
balance which makes the fidelity increase and remain constant with
values above 99.5\%. On the top of that, the exploration range goes
progressively to zero. The ideal experiment is an excellent example of how  quickly the algorithm
could reach a high fidelity above 99.9\% and also guarantee  the
convergence of $\Delta\rightarrow0$. In this way, once the exploration
range has become so small, it is assured that the agent does not go
to another state. In other words, it is proved that the agent has
definitely converged to the environment to a large fidelity.

In Table \ref{tab:Value-of-fidelity-Delta--0}, we sort some results
of the experiments run in Rigetti 8Q-Agave cloud quantum computer
from larger to smaller value of the fidelity. As we can see, $\mathcal{F}$
is close to 100\% in most of the experiments when the exploration
ratio $\Delta$ is approaching zero. Thus, we are succeeding in
adapting the agent to the environment state. From  these  data, we can
also draw the conclusion that the convergence of the agent to the
environment state is guaranteed whenever $\Delta\rightarrow0$. We
did not find any case where the exploration range is close to zero
and the fidelity below 90\%.

\begin{table}[H]
\centering
\caption{Value of the fidelity for $\Delta\rightarrow0$.\label{tab:Value-of-fidelity-Delta--0}}
\begin{tabular}{cccccccc}
\toprule
\boldmath$\Delta$ & \textbf{0.36} & \textbf{0.24} & \textbf{0.18} &\textbf{ 0.03} & \textbf{0.05} & \textbf{0.24} & \textbf{0.16}\tabularnewline
\midrule
$\mathcal{F}\,(\%)$ & 99.89 & 99.72 & 99.53 & 99.20 & 97.72 & 97.53 & 94.72\tabularnewline
\midrule 
Initial environment state & $|\varepsilon_{6}\rangle$ & $|\varepsilon_{2}\rangle$ & $|\varepsilon_{1}\rangle$ & $|\varepsilon_{3}\rangle$ & $|\varepsilon_{4}\rangle$ & $|\varepsilon_{1}\rangle$ & $|\varepsilon_{3}\rangle$ \tabularnewline
\bottomrule 
\end{tabular}
\end{table}
\unskip

\section{Discussion\label{part:Conclusion}}

In this  study, we   performed the implementation of the measurement-based
adaptation protocol with quantum reinforcement learning of     \cite{Measurement based adaptation protocol}. Consistently with it,
we   checked that indeed the fidelity increases reaching values
over 90\% with fewer than about 30 iterations in real experiments. We
did not observe any case where $\Delta\rightarrow0$ and $\mathcal{F}<90\%$.
Thus, to a large extent, the protocol succeeds in making the agent
converge to the environment. If we wanted to apply this algorithm
to any subroutine, it would be possible to track the evolution of
the exploration range and deduce from it the convergence of the agent
to the environment. This is because the behaviour of $\Delta$ has
proven to be closely related to the fidelity performance.

We can conclude that there is still a long
way to be travelled until the second quantum revolution gives rise
to well-established techniques of quantum machine learning in the lab. However,
this work is encouraging since it sows the seeds for turning quantum
reinforcement learning into a reality,     and the future implementation of semiautonomous quantum agents. 

We point out that another implementation of   \cite{Measurement based adaptation protocol} in a different platform, namely  quantum photonics, has been recently achieved, in parallel to this work \cite{Hefei}. The quantum photonics platform is more suited to quantum communication, while the current one, superconducting circuits, is more suited to quantum computation. On the other hand, there are interesting efforts for building quantum computers based on quantum photonics that exploit quantum machine learning techniques \cite{Xanadu}. The importance of these two experiments of the proposal in   \cite{Measurement based adaptation protocol} is the realisation that this proposal works well irrespective of the quantum platform in which it is implemented. Both experiments are complementary and it is valuable to show their good performance for different quantum technologies.
The impact of quantum machine learning,     and in particular, quantum reinforcement learning, in industry could be large in the short to mid term,     and companies such as Xanadu \cite{Xanadu}, D-Wave \cite{DWave}, IBM \cite{IBMQExperience}      and Rigetti \cite{Rigetti} are strongly investing on these efforts.
\vspace{6pt}
\newpage
\authorcontributions{J.O.-S. wrote the code and performed the experiments and simulations with Rigetti Forest. J.O.-S., J.C., E.S.      and L. L. designed the protocol, analysed the results, and wrote the manuscript. All authors have read and agree to the published version of the manuscript.}

\funding{J. C. acknowledges the support from Juan de la Cierva grant IJCI-2016-29681. We also acknowledge funding from Spanish Government PGC2018-095113-B-I00 (MCIU/AEI/FEDER, UE)      and Basque Government IT986-16. This material is also based upon work supported by the projects OpenSuperQ and QMiCS of the EU Flagship on Quantum Technologies, FET Open Quromorphic, Shanghai STCSM (Grant No. 2019SHZDZX01-ZX04)      and by the U.S. Department of Energy, Office of Science, Office of Advanced Scientific Computing Research (ASCR) quantum algorithm teams program, under field work proposal number ERKJ333.}

\acknowledgments{The authors acknowledge the use of Rigetti Forest for this work. The views expressed are those of the authors and do not reflect the official policy or position of Rigetti, or the Rigetti team. We thank I. Egusquiza for useful discussions.}

\conflictsofinterest{The authors declare no competing interests.}


\reftitle{References}

\end{document}